\documentstyle[12pt,epsf]{article}

\newlength{\extralineskip}
\newcommand{\junk}[1]{}

\newcommand{\R}{{\rm l\!R}}
\newcommand{\beq}{\begin{equation}}
\newcommand{\eeq}{\end{equation}}
\newcommand{\bea}{\begin{eqnarray}}
\newcommand{\eea}{\end{eqnarray}}
\newcommand{\nn}{\nonumber}

\newcommand{\A}{{\cal A}}
\newcommand{\B}{{\cal B}}
\newcommand{\F}{{\cal F}}
\newcommand{\D}{{\cal D}}

\renewcommand{\a}{{\bf a}}
\renewcommand{\d}{{\bf d}}

\newcommand{\Tr}{{\rm Tr}}

\begin{document}
\thispagestyle{empty}
\begin{center}
{\Large{\bf Topological Holography}}
\end{center}

\vskip 10mm
\centerline{Viqar Husain and Sebastian Jaimungal\footnote{emails: husain@physics.ubc.ca, jaimung@physics.ubc.ca}}
\begin{center}
Department of Physics and Astronomy \\
University of British Columbia \\
6224 Agricultural Road \\
Vancouver, British Columbia V6T 1Z1, Canada.
\end{center}
\vskip 10mm
\begin{abstract}
\noindent
We study a topological field theory in four dimensions on a manifold
with boundary. A bulk-boundary interaction is introduced through a novel
variational principle rather than explicitly. Through this scheme we find 
that the boundary values of the bulk fields act as external sources for 
the boundary theory. Furthermore, the full quantum states of the theory 
factorize into a single bulk state and an infinite number of boundary 
states labeled by loops on the spatial boundary. In this sense the theory 
is purely holographic. We show that this theory is dual to Chern-Simons 
theory with an external source. We also point out that the holographic 
hypothesis must be supplemented by additional assumptions in order to 
take into account bulk topological degrees freedom, since these are 
apriori invisible to local boundary fields.
 \end{abstract}

\newpage\setcounter{page}{1}

There has been much recent interest in the interplay between bulk and
boundary dynamics. The two main directions being explored presently
are (i) the Maldacena conjecture\cite{Ma97}, which postulates a
relationship between a bulk string/M-theory and a boundary conformal
field theory (also known as the AdS/CFT correspondence), and (ii) the 
holographic hypothesis \cite{t'H,Thorn,Su95}, which states that all 
information about a theory in the bulk of a bounded region is available, 
in some sense, on the boundary of the region. In particular, the AdS/CFT 
correspondence has been viewed as an example of the holographic 
hypothesis \cite{suswit}. 

The first of these directions is based in part on the observation that
the symmetry group of $d+1$-dimensional anti-deSitter space-time 
$SO(2,d)$ is the same as the conformal group of Minkowski space-time in 
$d$ dimensions. More specifically, a statement of the conjecture  
\footnote{ There is a more general, and fully quantum mechanical statement 
of this conjecture, where the left hand side includes functional 
integrals over bulk fields $\phi_i$ which have boundary values 
$\phi_i^B$, and over asymptotically anti-deSitter metrics. The statement 
of the correspondence given above is effectively the  tree level evaluation 
of the left hand side, and represents all its tests to date!}
is \cite{Wi98,gkp} 
\beq
  {\rm exp} \left[- \int_{AdS_{d+1}} {\cal L}_{\rm SUGRA}
(\phi_i(\phi_i^B)) \right] 
= \left< {\rm exp} \int_{\partial AdS_{d+1}} {\cal O}^i\phi_i^B \right>_{\rm
CFT}. 
\eeq
The left hand side of this equation is the  evaluation of the 
Euclidean supergravity action on the {\it classical}  solutions for 
which the background is the (d+1)-dimensional anti-deSitter metric. 
$\phi_i$ represent the bulk dynamical fields in the solution. 
The surface integral in the supergravity action, which is a functional 
of the boundary value $\phi_i^B$ of $\phi_i$, is a crucial input here. 
The right hand 
side is the {\it quantum} expectation value of the primary field ${\cal O}^i$ 
of some conformal field theory on the boundary of AdS, where the boundary 
value of the bulk field $\phi_i^B$ acts as an external source. Thus, 
this conjectured equality provides a way of computing conformal field 
theory correlation functions from classical supergravity. It therefore 
provides a classical-quantum duality for a sector of the solution space of 
supergravity, (-- the sector for which the metric is anti-deSitter). 
A key feature of this prescription is that a classical bulk field 
provides, via its boundary value, an external source for a boundary 
quantum theory.  This feature appears in the model we discuss below. 
 
The second direction in this bulk-boundary interplay is (at least
partly) motivated by arguments concerning black holes: The fact that
the entropy of a black hole is proportional to its area suggests the
possibility that the theory describing microstates of a black hole is
either (i) a surface theory, or, (ii) a bulk theory whose states are
``visible'' on the bounding surface in such a way that the entropy
becomes proportional to the surface area. This is closely connected to
and motivated by the Beckenstein bound argument \cite{beck,ls}.

There are in fact (at least) four possible definitions of what  
holography may mean: 
\def\theenumi{\roman{enumi}}
\def\labelenumi{(\theenumi)}
\begin{enumerate}
\item For a theory defined in a bounded spatial region, all bulk degrees 
of freedom are ``visible'' on the boundary of the spatial region through 
a physical process. This may be done via a ``screen mapping'' 
\cite{Su95,CJ}.

\item For an $n$-dimensional theory, T1, containing bulk and boundary 
degrees of freedom, there is an $(n-1)$-dimensional theory, T2, which 
captures all the degrees of freedom of T1. Then the possibility is open 
that T2 is itself defined on a manifold with boundary, having both bulk 
and surface degrees of freedom. This is obviously different from (i).

\item The same as (ii) with the extra condition that T2 is a theory 
defined strictly on the boundary of the region on which T1 is defined.
In this case, T2 has only bulk degrees of freedom, since the boundary
of a boundary vanishes.

\item All the degrees of freedom of a theory in a bounded region are 
associated with its boundary. In this case holography is automatic. 
\end{enumerate}

The AdS/CFT correspondence appears to fall in category (iii). 
However, the Beckenstein bound argument, which requires entropy 
to be proportional to the bounding area, appears to be consistent 
with all four possibilities. 

\begin{figure}
\epsfxsize=75mm
\hspace{55mm}\epsffile{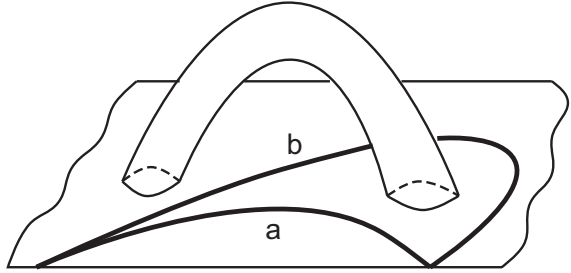}
\caption{Two Wilson lines piercing the boundary at the same points
where path (a) is trivial while (b) wraps around a  handle.}
\label{WilsonLine}
\end{figure}

An interesting possibility which should be taken into account in a
definition of holography is the case of theories which have bulk
topological degrees of freedom, associated for example with
handles. It is then possible that these are not visible on the
boundary. An example is a Wilson line observable with
end points on the boundary which may or may not wrap around a bulk
handle (see Figure \ref{WilsonLine}). Then only (ii) seems viable as
a definition of holography, and may exclude the AdS/CFT case (iii).

In this paper we describe a theory which is holographic in the sense
of both (ii) and (iv) above. The quantum states factorize into bulk and 
boundary states, 
with a unique bulk state. It has the unusual property that 
all its quantum states are effectively associated with loops lying on 
the spatial boundary. The dynamics of the loop states is trivial, so their
worldlines are cylindrical. Furthermore, the bulk states give rise to 
external sources for the boundary quantum theory in a natural manner. 
This latter feature is similar to one of the key features in the AdS/CFT
correspondence, albeit in a simpler context. Models of this type may
be used to probe the limitations of the holographic hypothesis and perhaps 
the AdS/CFT correspondence. One such limit, as briefly mentioned above,
appears to be that bulk topological degrees of freedom are not
captured by the values of local fields on boundary.

{\it The  Model:}  The theory we consider is a topological field 
theory on a four-dimensional manifold  $\R\times\Sigma^3$, defined by 
an unusual specification of the variational principle.  The action is 
\beq
S[\A,\B;\a] = 
\int_{\R\times\Sigma^3}~\Tr\left\{\B\wedge \F(\A) 
+ \frac \Lambda 2\ ~\B \wedge 
\B \right\} + k\int_{\R\times\Sigma^2}~\Tr\left\{\a\wedge \d \a 
+ \frac 2 3 \ \a
\wedge \a \wedge \a \right\}
\eeq
where $\B$ is a Lie-algebra valued two form, $\A$ and $\a$ are
Lie-algebra valued one forms, and $\F(\A)$ denotes the curvature
two-form. $\Sigma^2$ is the 2-boundary of the ``spatial'' surface 
$\Sigma^3$. 

This action contains no explicit interaction terms between the bulk and 
boundary fields. However, the action alone does not determine the 
equations of motion or the subsequent canonical structure, since it must 
be supplemented by a variational principle.  Particular choices of the 
variational principle can lead to a situation in which the bulk BF-theory 
is coupled to the boundary Chern-Simons theory.  We will invoke the 
particular scheme where the field $\a$ must be varied in accord with 
the variations of the field $\A$ on the boundary.

The variation of this action is   
\beq
\delta  S =  \Tr \int_{\R\times\Sigma^3} \left[ 
 \delta \B \wedge ( \F(\A) + \Lambda \B )
+ \delta \A \wedge (  \D_{\A} \B ) \right] 
+ \Tr \int_{\R\times\Sigma^2}\left[ k~\F(\a) \wedge \delta \a 
+ B \wedge \delta\A \right],  
\eeq
the variational principle is well-defined if we require that  
\bea
\delta \a &=& \delta \A|_{bd.} \nn\\
\delta_\A S &\equiv& S[\A+\delta \A, \B; \a+\delta \A|_{bd.}] 
- S[\A, \B; \a], 
\label{interact}
\eea
and the usual requirement that all surface terms in the variation 
vanish. The constraints on the variations of the fields may be viewed as
giving rise to the equations of motion for the boundary theory.
Ordinarily, a variational principle is supplemented by conditions such 
as the vanishing of the variations of certain fields on the boundary. Our
prescription is unusual only in that it fixes the variation of certain
fields to equal certain other variations on the boundary.
 
A question concerning our approach so far is why we do not, perhaps
more simply, consider the above action with $\a = \A$ at the
outset. The reason is that doing this gives a different theory: 
functional differentiability requires that the fields $\A$ and $\B$ 
satisfy the condition $\F(\A) = \B $ on the boundary. In our case on 
the other hand, the bulk field $\B$ provides a source for the 
{\it independent} boundary curvature $\F(\a)$.

The phase space variables are identified by writing 
 \bea
S[A,A_0, B, B_0; a,a_0] &=& 
\int dt \left[ ~
\int_{\Sigma^3}~\Tr\left\{ B \wedge {\dot d} A + B_0 \wedge
\left( F(A)+\Lambda~ B \right) 
+ A_0 \wedge D_A B \right\} \right.\nn\\
&& \left. \hspace{5mm}
+~k \int_{\Sigma^2}~\Tr\left\{ a\wedge {\dot d} a 
+ a_0 \wedge F(a) \right\}
+ \int_{\Sigma^2} \Tr ~A_0 \wedge B
\right]\label{Haction}
\eea
In this expression we have written $\A=A_0+A$, $\B=B_0+B$ and
$\a=a_0+a$, where the fields carrying a subscript $0$ contain the
`time' components, and the other fields contain the spatial components
of the respective forms; $\dot d$ represents the time component
of the exterior derivative; $F(\cdot)$ is the spatial part of the
curvature two form; and $D_A$ is the spatial part of the covariant
derivative. The boundary contribution has two terms: the first is the
Chern-Simons action in Hamiltonian form, and the second is from 
an integration by parts in the bulk part of the action. This latter
term provides a bulk source for the boundary curvature.

The canonical structure of the theory is obvious from  
(\ref{Haction}): $A_a^i$ and $E^{ai}\equiv \epsilon^{abc} B_{bc}^i$
are the canonically conjugate variables in the bulk, $a_1^i$ and
$a_2^i$ are canonically conjugate on the
boundary; ($a,b,\cdots $ are spatial indices $1,2,3$ in the bulk, and
$1,2$ on the boundary;  $i,j,\cdots $ are Lie algebra indices). The
time component fields ($a_0, A_0$ and $B_0$) appear as Lagrange
multipliers, and varying these fields gives the phase space constraints.
Since $\Sigma^3$ has a boundary, there is an additional boundary 
constraint arising from functional differentiability of the action
(recall that $a_0$ must be varied with $A_0$ (\ref{interact})). The 
relevant variations are
\bea
\delta_{A_0} S &=& \int dt \left\{
\int_{\Sigma^3} \Tr~\left\{ \delta A_0 \wedge D_A B \right\}
+\int_{\Sigma^2} \Tr~\left\{ \delta A_0 \wedge \left( B + k~F(a) \right) 
\right\}\right\} \label{A0var}\\
\delta_{B_0} S &=& \int dt \int_{\Sigma^3} \Tr~\left\{ \delta B_0 \wedge
\left(F(A) + \Lambda~B\right) \right\}
\eea

>From the above it is clear that the Hamiltonian is a linear combination 
of constraints. The bulk constraints are
\beq
G^i \equiv\left.(D_a E^{ai})\right|_{\Sigma^3} = 0
~,\qquad
J^{ai} \equiv \left.\left(\epsilon^{abc}F(A)_{bc}^i 
+ \Lambda~E^{ai}\right) \right|_{\Sigma^3}=0
\eeq
 In addition to these there is a {\it surface} constraint 
 \beq
H^i \equiv \left.\left( E^{3i} +
k~\epsilon^{ab}F(a)_{ab}^i \right)\right|_{\Sigma^2} = 0,
\label{FuncDiff}
\eeq
due to the presence of the surface integral in (\ref{A0var}).  

It is clear that both the bulk and boundary constraints are first
class and provide the complete prescription for classical Hamiltonian
evolution. The bulk phase space variable $E^{3i}$ provides a source
for the boundary curvature, and the boundary fields $\a$ evolve via
the first class constraint (\ref{FuncDiff}). This evolution is a gauge
transformation on $\a$. As a result the bulk evolution of $(A,E)$ is
consistent with the boundary evolution of $\a$. To see this more
explicitly, start from an initial classical configuration where
$E^{3i}$ fixes $F(\a)$. We must now ensure that the evolved $E^{3i}$
and $F(\a)$ also satisfy the constraint (\ref{FuncDiff}). That this is
indeed the case is ensured by our variational principle, and is
easiest to see directly in the covariant picture.

{\it Observables:} The Hamiltonian of the bulk theory is a linear 
combination of first class constraints. Therefore the gauge invariant 
observables are phase space functionals that have weakly vanishing 
Poisson brackets with the constraints. For $\Lambda\ne 0$ we have not 
been able to find any observables. While we do not have a proof of this,
\footnote{A proof may be devised along the following lines: Write 
down all the basic local and non-local Gauss law invariant variables. 
The local ones are  combinations of the electric and magnetic fields 
with internal indices contracted; there are four such variables. 
The non-local ones are the traces of electric field insertions between 
holonomy segments such as 
$Tr[E^a(x_0)U_\gamma(x_0,x_1)E^b(x_1)U_\gamma(x_2,x_3)E^c(x_3)\cdots]$; 
these are a countably infinite set. (In the limit 
of the loop $\gamma$ shrinking to a point, these become functions of the 
local variables). Consider the general Gauss law invariant function to be 
an arbitrary function of these variables, and calculate its Poisson 
bracket with $J^a$, and see if the result can be made to vanish.}
the fact that there is a unique solution of the bulk Dirac quantization 
constraint (described below) suggests there are no bulk observables. 
On the other hand, for $\Lambda = 0$ there are two types of bulk 
observables. One type is parametrized by loops (and are traces of  
holonomies), while the other is parametrized by both loops and 
surfaces \cite{VH}. In this paper we will be interested in the case of
non-vanishing $\Lambda$.

Contrasted with the bulk case, there are  an infinite set of boundary 
observables for non-zero values of $\Lambda$. Since the boundary 
constraint generates Yang-Mills gauge transformations on $\a$, the 
boundary observables are traces of the holonomy of $\a$ for {\it all} 
loops lying on the spatial boundary $\Sigma^2$. Denoting these 
observables by 
\beq
T_\gamma[\a] \equiv {\rm Tr\ P~exp}\int_\gamma \a
\eeq
for loops $\gamma$, their Poisson algebra is 
\beq 
\{ T_\alpha[\a], T_\beta[\a] \} = 
\Delta(\alpha,\beta) \left( T_{\alpha\circ\beta}[\a] 
- T_{\alpha\circ\beta^{-1}}[\a]\right),
\eeq
where 
\beq
\Delta(\alpha,\beta) 
= \int ds\int dt ~\epsilon_{ab}~\dot{\alpha}^a(s)~\dot{\beta}^b(t)~
\delta^2(\alpha(s)-\beta(t))
\eeq
measures a weighted intersection number of the loops $\alpha$ and $\beta$, 
and $\beta^{-1}$ denotes traversal of the loop in the opposite sense
($\dot{\alpha}^a(s)$ is the tangent vector to the loop at the parameter 
value $s$).

Consequently, the 4-dimensional theory we have outlined has an
infinite number of boundary observables parameterized by loops lying
in the 2-boundary $\Sigma^2$ of $\Sigma^3$. The observables form a
closed infinite dimensional Poisson algebra.

On the constraint surface, the bulk contains no local degrees of
freedom: for gauge group $SU(N)$ there are $3(N^2-1)$ configuration
variables $A_a^i$ and $4(N^2-1)$ first class bulk constraints. This
means that there can be at most a finite number of bulk topological
degrees of freedom. However, since there appear to be no bulk
observables to evaluate on the constraint surface, it is likely that
there are {\it no} bulk degrees of freedom in this theory.  This
means that the large gauge freedom may be used to set all bulk fields
to zero.  Nevertheless there may still be an infinite number of
surface observables on the reduced phase space. To see this take
$E^{3i}$ to be zero everywhere in the bulk but keep it arbitrary and
non-zero on the boundary. This gives the curvature of $\a$ via the
surface constraint (\ref{FuncDiff}). Conversely, given {\it any}
boundary field $\a$, the boundary field $E^{3i}$ is determined. This 
field may then be arbitrarily extended into the interior. A particularly 
simple case is where $E^{3i}$ vanishes on the boundary. Then the reduced
phase space of our four-dimensional theory is the (finite dimensional)
moduli space of flat connections on the 2-boundary $\Sigma^2$, which
may be a surface of arbitrary genus.

The boundary observable algebra given here is reminiscent of, but
fundamentally different from, the construction that gives the
Kac-Moody boundary observable algebra associated with 3-dimensional
gravity. However, the Brown-Henneaux\cite{BH} construction of observables 
for the latter theory is intrinsically  dependent on the fall-off
conditions of the bulk fields.  In particular, almost all the
Brown-Henneaux observables vanish identically on solutions of 3d
gravity such as the BTZ black hole \cite{BTZ}.  In our construction
this is manifestly not the case -- the fields in the bulk may be
obtained from {\it any} connection $\a$ (on $\Sigma^2$), whose
curvature gives the boundary value of $E^{3i}$ via the boundary
constraint (\ref{FuncDiff}). Conversely, given bulk fields, the
boundary value of $E^{3i}$ fixes the curvature of the boundary
connection $\a$.  As such, all the observables are non-zero on 
generic solutions, unlike the case of 3-dimensional gravity.

{\it Quantization:} 
We will carry out a quantization in the Hamiltonian formulation described 
above. There are two ways to approach this: (i) convert the classical 
constraints into operator equations in a suitable representation and  
attempt to solve them for the quantum states, or (ii) find a 
representation of the algebra of classical gauge invariant observables. 

In a model such as the one described here, it is possible to carry out
a ``hybrid'' quantization using both of these approaches
simultaneously.  This is because the bulk and boundary states have a
natural separation.  Specifically, the bulk constraint can be imposed
as a Dirac quantization condition (since we do not have any bulk
observables to find a representation of), while the boundary sector
can be quantized by finding a representation of the algebra of the
$T_\alpha(\a)$ observables. We will follow this procedure, and define
the bulk quantum constraint as acting by the identity on boundary
states.

Consider first the bulk constraints and use the connection
representation; $A_i$ are treated as configuration variables, and
their conjugate momenta $E^i$ are treated as functional derivative
operators
\beq
E^i\to -i\frac{\delta}{\delta A_i}.
\eeq 
 We assume that the quantum states may be written as the 
product 
$$
\Psi = \psi_{\Sigma^3}(\A)\otimes \psi_{\Sigma^2}(\alpha),
$$ where $\alpha$ denotes the parameterization of boundary states (to
be discussed below).  The bulk constraint $J^{ai}$ gives the condition
\beq
\left\{ \left.\left(\epsilon^{abc} F(A)^i_{bc} 
-i\Lambda\frac{\delta}{\delta A_a^i} \right)\right|_{\Sigma^3} 
\psi_{\Sigma^3}   \right\}
\otimes \left\{ I|_{\Sigma^2} \ \psi_{\Sigma^2}(\alpha) \right\}
= 0 \label{QmBkCon}
\eeq
where $I$ is the boundary identity operator. The unique solution of
this constraint is,
\beq
\psi_{\Sigma^3}(A) \equiv \exp\left\{ -\frac{i}{\Lambda} \int_{\Sigma^3}
\Tr~\left\{ A\wedge dA + \frac{2}{3} A\wedge A\wedge A \right\} 
-\frac i {2 \Lambda} \int_{\Sigma^2} \Tr~(A_1~ A_2) \right \} \label{BkQmSt}
\eeq
This state also satisfies the bulk Gauss constraint. The surface term
in the exponential is necessary to guarantee that the functional
derivative gives $\epsilon^{abc}F_{bc}^i$, and does not spoil the bulk
Gauss law invariance. The solution in the case where $\Sigma^3$ is
compact without boundary is the bulk part of this functional, and has
been discussed in \cite{GH}.

This Chern-Simons state is not directly related to the Chern-Simons part 
of the original action (\ref{Haction}): the state is still a solution 
of the bulk constraints if $\Sigma^3$ has no boundary.  
Furthermore, although (\ref{BkQmSt}) is a solution of the bulk constraint, 
the functional does not transform trivially under the generators of the 
constraint (i.e. the Poisson bracket of the constraint with the  
functional does not vanish). This is unlike the Wilson loop functional, 
which is simultaneously a Gauss law invariant classical observable, as 
well as a quantum state satisfying the quantum Gauss constraint. 

(The latter result might seem surprising at first, and in apparent
violation of the intuition derived from the Gauss law. However, it is
also illustrated in a simple quantum mechanical example\footnote{The
authors would like to thank W. Unruh for pointing out this
example.}. Consider the constraint equation, $$ (\hat{x}+\alpha
\hat{p})\psi = 0 $$ where $\alpha$ is some dimensionful constant. The
solution to this constraint in the $x$ representation is $$
\psi \propto \exp \left\{-\frac{i}{2\alpha} x^2 \right\}
$$ 
This function is clearly not invariant under the transformation 
 $x\rightarrow x + \alpha$ generated by the constraint. 
However, the canonical transformation $\tilde p = x+ \alpha p, \tilde x 
= x/\alpha$ reduces the constraint to ${\tilde p} {\tilde \psi} = 0$ 
whose solution is $\tilde \psi(\tilde x) = {\rm constant}$, which 
does commute with constraint.)

Turning now to the boundary dynamics, we choose to quantize this sector 
by finding a representation of the algebra of the boundary observables 
$T_\alpha[\a]$. This is easiest to do in the loop representation, and we 
follow here the prescription used in the approach to non-perturbative 
quantum gravity \cite{RS, AI}. The holonomy observables $\hat{T}_\alpha$ are 
defined to act on loop states $|\beta>$ by 
\beq
    \hat{T}_\alpha |\beta> := i\hbar \Delta(\alpha,\beta)
\left( |\alpha \circ \beta> - |\alpha\circ \beta^{-1} > \right).
\eeq 
>From this definition it follows that
\beq
[\hat{T}_\alpha, \hat{T}_\beta ] = \Delta(\alpha,\beta) 
\left( \hat{T}_{\alpha\circ\beta} 
- \hat{T}_{\alpha\circ \beta^{-1}} \right).
\eeq

Thus, in this approach the boundary states are the loop kets $|\alpha >$, 
and the full quantum state of the theory is the product 
\beq
   |A,\alpha> = \psi[A]|\alpha>. 
\eeq

The loop states $|\alpha>$ are not all independent: the states are 
traces of holonomies in the connection representation and are subject 
to the Mandelstam identities induced by the trace relations on $SU(N)$ 
matrices. Furthermore, because there is a unique bulk state, the labeling 
of quantum states is effectively only by loops. An inner product on this 
space of states may be defined as $<\alpha|\beta> = \delta_{\alpha\beta}$. 
This completes the description of the quantum theory.

{\it Discussion:} The model we have described has a number of unusual
features which are useful to compare with Chern-Simons theory on a
manifold with boundary, and with 2+1 gravity in particular. These
latter theories have the property that they are topological in the bulk 
and, with particular fall-off conditions on the fields, induce a Kac-Moody
algebra of observables (which are all constants of motion) on the
boundary. These theories thus have non-trivial bulk and boundary 
observables, (if the bulk has non-trivial topology).  
The boundary observables may be viewed as the observables
of a two-dimensional boundary conformal field theory. Apparently for this 
reason, these theories have been viewed as an example of the AdS/CFT 
correspondence \cite{hennetal}, and therefore an example of holography. 
However this does not correspond to holography for any of the possible 
definitions given above. What would be required for a correspondence 
with one of these definitions is the specification of a 2-dimensional 
theory that has both the bulk and boundary observables of 3-dimensional 
gravity. (See \cite{hyun} in this regard.) 

For the case of our 4-dimensional model, the 3-dimensional theory that
has the {\it same} observables algebra is Chern-Simons theory coupled
to an external source $J^a$ (which plays the role of $E^{3i}$), ie. the
action is the Chern-Simons one with the additional term
$\frac{1}{\Lambda}\int_{R\times\Sigma^2} A_aJ^a$.  Consequently, in addition 
to viewing
our model as an example of type (iv) holography, we can also view it as 
type  (iii) holography.

In summary, we have discussed some aspects of holography in a
4-dimensional model in which all degrees of freedom are associated
with loops on a 2-dimensional boundary. We have: (i) pointed out that 
topological bulk observables are missed by any boundary theory that is
directly induced by the bulk fields, (ii) suggested that this shortcoming 
may be side-stepped by broadening sufficiently the definition of holography,
and (iii) given a 3-dimensional theory that has the same observable
algebra as our 4-dimensional model. This provides a concrete example of
duality: theories in different spacetime dimensions having the same 
classical and quantum observable algebra.

This work was supported by the Natural Science and Engineering Research 
Council of Canada, and by a University of British Columbia Graduate 
Fellowship. We thank G. W. Semenoff and W. G. Unruh for discussions.

\end{document}